\documentclass[conference]{IEEEtran}
\IEEEoverridecommandlockouts

\usepackage{cite}
\usepackage{amsmath,amssymb,amsfonts}
\usepackage{algorithmic}
\usepackage{graphicx}
\usepackage{graphics}       
\usepackage{epsfig}
\usepackage{textcomp}
\usepackage{xcolor}
\def\BibTeX{{\rm B\kern-.05em{\sc i\kern-.025em b}\kern-.08em
    T\kern-.1667em\lower.7ex\hbox{E}\kern-.125emX}}

\usepackage{lipsum}
\usepackage{mathtools, cuted}

\usepackage{subcaption}
\usepackage{import}
\usepackage{svg}
\usepackage{mathtools} 

\begin{document}

\title{Study of Frictional and Impact Transients in Active-Passive Mechanical Pair}



\author{Michael Ruderman and Francesco De Rito  
\thanks{M. Ruderman is with Department of Engineering Sciences, University of Agder (UiA).
Postal address: P.B. 422, Kristiansand, 4604, Norway. \newline
Corresponding author: email {\tt\small michael.ruderman@uia.no}}
\thanks{F. De Rito is with University of Padova, Vicenza, Italy.
}
\thanks{This work was partially supported by the ERASMUS+ program. \newline}
\newline
\thanks{\textcolor[rgb]{0.00,0.00,1.00}{AUTHORS MANUSCRIPT ACCEPTED TO IEEE ICM2025}}
}

\maketitle

\bstctlcite{references:BSTcontrol}

\begin{abstract}
We consider an active-passive mechanical pair in which the
relative motion of the latter is constrained by the mechanical
impact. The system dynamics is described by the previously
introduced modeling frameworks of force transition and dissipation
through the nonlinear Coulomb friction and structural damping, the
later in accord with Hertzian contact theory. The focus of the
recent study is on combining both interaction mechanisms, and the
detailed experimental evaluation which discloses validity of the
modeling assumptions. Such mechanical pair interactions can be
found in various mechatronic systems and mechanisms, like for
example clutches, backlash elements, sliding items on the shaking
and inclining surfaces, conveyor belts and others. This practical
study demonstrates and discusses the transients of a vibro-impact
dynamics and shows theoretical developments in line with
experimental evaluation.
\end{abstract}


\section{Introduction}  
\label{sec:1}

Multiple mechatronic systems and mechanisms involve nonsmooth
dynamics (see e.g. \cite{brogliato2016} for basics) with impacts,
in addition to kinetic friction (see e.g.
\cite{ruderman2023analysis} for introduction) of the moving parts
in contact with each other. Both well-known mechanical effects are
often unavoidable due to the structural properties and the
corresponding functionality of a mechanism at hand. Both also
involve the associated energy dissipation (mostly nonlinear in the
nature) and influence (in more peculiar way) the transient
behavior of each mechanical pair. One of the clearest examples of
the combined effect of impact and friction dynamics is backlash
\cite{goodman1963}, also known as mechanical play. Hybrid
approaches are known for modeling and controlling backlash, e.g.
\cite{rostalski2007}, and detecting and identifying it
\cite{ruderman2019,Ruder21}. Apart from the backlash, the
vibro-impact applications can be found, for instance, in clutches,
forging and riveting machines, pneumatic and hydraulic hammers and
drills (cf. e.g. \cite{babitsky98} for basics), but also in
shaking and tilting surfaces, conveyor belts and more. Later, the
impact dynamics with restitution \cite{hunt1975} was found useful
even in the robotics, for simulating contacts with environment
e.g. \cite{marhefka1996simulation}, or by attempting to describe
the walking gait of walking robots, cf. e.g.
\cite{freidovich2009}.

Systems which include nonsmooth mechanics are understood to have
hybrid dynamics \cite{goebel2009}, with the corresponding guard
and jump maps in the state space, and (eventually) differential
inclusions instead of differential equations where necessary. Such
systems become hybrid in terms of continuous and switched
solutions (correspondingly state trajectories) and require
particular attention in every respect when it comes to modeling,
identification, and control. For a tutorial on nonsmooth analysis
and stability we refer to e.g. \cite{cortes2008}. For basics on
hybrid and switched systems, see also \cite{liberzon2003}.

In this paper, two modeling approaches are combined together for
describing and experimentally studying the frictional and impact
transients in an active-passive mechanical pair. The first
modeling framework, introduced in \cite{ruderman2022dynamics},
formalizes the dynamic interactions of two inertial bodies that
are connected to each other exclusively via the friction surface
and governed by the nonlinear Coulomb friction. The second
modeling approach \cite{ruderman2021stiffness} originates from
describing the contact force and corresponding structural damping
of the impact in a backlash pair, while using the restitution and
damping laws formulated in \cite{hunt1975}. Dynamic equations of
both modelings are combined in a hybrid setting, and share the
same state variables and switching conditions. One of the cores of
this study is a series of experiments designed for the dedicated
laboratory setup to reveal meaningful motion transitions.

The rest of the paper is organized as follows. In section
\ref{sec:2}, we provide the necessary preliminaries of the overall
system modeling, following the original works
\cite{ruderman2022dynamics,ruderman2021stiffness}. Section
\ref{sec:3} describes the used tribological setup, cf.
\cite{ruderman2023robust}, specially developed for frictional and
impact experiments with one active and one passive moving body in
contact. The comparison between the modeled and measured system
response is reported in section \ref{sec:4}. The paper is briefly
concluded by section \ref{sec:5}.

\section{Preliminaries}  
\label{sec:2}

\begin{figure}[!h]
\centering
\includegraphics[width=0.7\columnwidth]{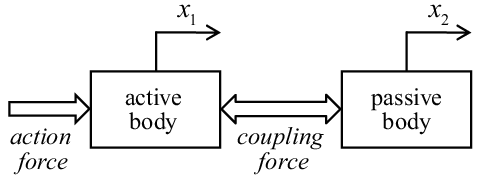}
\caption{Action and reaction flow in active-passive pair.}
\label{fig:pair}
\end{figure}

In this section, we first recall both modeling frameworks
\cite{ruderman2021stiffness} and \cite{ruderman2022dynamics} used
for describing the frictional and impact transients in an
active-passive mechanical pair. A principal structure of an
active-passive mechanical pair is shown in Fig. \ref{fig:pair},
with the corresponding action and reaction flow.

\subsection{Coupling via frictional interface}  
\label{sec:2:sub:1}

Considering a mechanical pair consisting of an active and a
passive subsystem, the relative motion of the first occurs in the
coordinates $x_1$ and that of the second in $x_2$, cf. Fig.
\ref{fig:pair}. Both are in the generalized coordinates (either
translational or rotational) with one degree of freedom and
governed by
\begin{equation}\label{eq:2:1}
\ddot{x}_1(t) = f_1\bigl(x_1(t), x_2(t), t \bigr), \quad
\ddot{x}_2(t) = f_2\bigl(x_2(t), x_1(t)\bigr).
\end{equation}
Note that the second subsystem in \eqref{eq:2:1} is
semi-autonomous and depends explicitly on the dynamic trajectories
of the first one. Assuming a flat and unconstrained contact of
both mechanical bodies of the subsystems given in \eqref{eq:2:1},
which is due to a normal load provided by the lumped mass $m_2$,
the generic modeling framework of the corresponding frictional
coupling was introduced in \cite{ruderman2022dynamics}. The
transient and steady-state behavior of the both moving bodies is
captured by
\newpage
\begin{strip}
\begin{eqnarray}
\label{eq:2:2}
  x_1-x_2 &=:& z, \\[2mm]
\label{eq:2:3}
  \Bigl(m_1 + m_2 \bigl(1-\bigl| \mathrm{sgn}(\dot{z}) \bigr| \bigr)  \Bigr) \, \ddot{x}_1 + a_1 \dot{x}_1
  + a_2 x_1 + b\,\mathrm{sgn}(\dot{z}) &=& u, \\
\label{eq:2:4}
   m_2 \ddot{x}_1 \Bigl( 1- \bigl|  \mathrm{sgn}(\dot{z})  \bigr|
   \Bigr) \frac{1}{2} \Bigl(1 - \mathrm{sgn} \bigl( |\ddot{x}_1| - b m_2^{-1} \bigr)
   \Bigr) - m_2 \ddot{x}_2 + b\,\mathrm{sgn}(\dot{z}) &=& 0.
\end{eqnarray}
\end{strip}
Here the dynamics equations \eqref{eq:2:2}--\eqref{eq:2:4}
correspond to the system \eqref{eq:2:1} for the case where the
first active body with the mass $m_1$ is feedback controlled,
while the coefficients $a_1, a_2 > 0$ accommodate both the state
feedback gains and (eventually) structural properties of the
active subsystem, like e.g. restoring spring and viscous damping.
Note that the state variables and signals in
\eqref{eq:2:2}--\eqref{eq:2:4} are without time argument for the
sake for brevity. The exogenous value $u(t)$ can be seen here as,
for example, a reference trajectory generator for the
feedback-controlled active subsystem \eqref{eq:2:3}. The coupling
between both subsystems occurs essentially due to the nonlinear
Coulomb friction with the coefficient $b > 0$. When the passive
body slips over the active one, the relative velocity between the
both is $\dot{z} \neq 0$. Otherwise, both subsystems are
considered to be in a coupled state with $\dot{z} = 0$, that
affects correspondingly the dynamics \eqref{eq:2:3},
\eqref{eq:2:4}. Important to notice is also that the classical
three-point-valued signum function is used here, i.e.
\begin{equation}\label{eq:2:5}
    \mathrm{sgn}(y)= \left\{%
\begin{array}{ll}
    1, & \; y>0, \\
    0, & \; y=0,\\
    -1, & \; y<0, \\
\end{array}%
\right.
\end{equation}
where an argument $y$ is the real number.

Further, it must be noted that if the active subsystem is robustly
controlled, i.e. the contact frictional force
$b\,\mathrm{sgn}(\dot{z})$ is not propagated back so as to affect
the motion trajectories $\bigl(x_1(t), \dot{x}_1(t) \bigr)$, then
the equation \eqref{eq:2:3} is substituted by
\begin{equation}\label{eq:2:6}
\ddot{x}_1(t) = f_1 ( t ).
\end{equation}
This is particularly the case we are going to consider in the
current study. For more details on the hybrid dynamics
\eqref{eq:2:2}--\eqref{eq:2:4}, an interested reader is further
referred to \cite{ruderman2022dynamics}.

\subsection{Vibro-impact dynamics}  
\label{sec:2:sub:2}

Upon contacting with mechanical motion limiters, the moving body
$m_2$ experiences impact transitions, which are associated with a
resulting repulsion force and structural damping. In case of
repetitive impact and separation, the so-called vibro-impact
system emerges, while the colliding body $m_2$ can be assumed as
absolutely stiff and the fixed frame as elastic, this for the sake
of simplicity and without loss of generality. Note that this
assumption is also in line with the general Hertzian theory of
non-adhesive elastic contacts.

The vibro-impact dynamics assumes $\dot{x}_2^o = - e \,
\dot{x}_2^i$, where the relative displacement rates are denoted by
the superscripts $\{i,o\}$, for indicating before (``in'') and,
correspondingly, after (``out'') the collision. Here we recall
that the restitution coefficient $e \in [0,\, 1]$ reduces the
relative velocity and, usually, is interpreted as a measure of the
degree of energy dissipation during an impact. Also we note that
the lower and upper boundaries of $e$ represent an absolute
plastic and, respectively, absolute elastic contact. For a limited
range of relatively low velocities, and for the most of materials
with a linear elastic range, it can be written with a tolerable
accuracy \cite{hunt1975}:
$$
e = 1 - \alpha \dot{x}_2^i.
$$
The coefficient $\alpha$ is characteristic for each considered
elastic material and geometry of the structure. During a
vibro-impact transition, the overall contact force is also
including the nonlinear structural damping, following
\cite{hunt1975}. Capturing the state of penetration of the body
$m_2$ into the elastic frame by
$$
p(t) = \int \limits_{t^i}^{t^o} \dot{x}_2(t) dt,
$$
where the integration limits $t^i$ and $t^o$ indicate the time
instants of the impact and separation, respectively, the contact
force is modeled as in \cite{ruderman2021stiffness} by
\begin{equation}
f(p) = \lambda p^n \dot{p} + k p^n. \label{eq:2:7}
\end{equation}
The coefficient $\lambda = 1.5 \alpha k$ is based on the energetic
balance of the restitution derived in \cite{hunt1975}. For the
construction materials (like e.g. steel or aluminium) $\alpha$ has
a relatively small value, thus keeping impact in a predominantly
elastic region. However, one can recognize a multiplicative
coupling between $\alpha$ and $k$, so that a practical
identification will have a trade-off between the used experimental
data and interpretation of the values range of the determined
parameters. The contact stiffness coefficient is $k > 0$, and the
structural damping increases with the depth of penetration, i.e.
with $\dot{p} > 0$. It is also worth recalling that for $n=3/2$,
the vibro-impact dynamics is consistent with the Herzian theory of
contacting spheres under static conditions, while for $n=1$ it
captures the most simple case of two flat surfaces, cf.
\cite{hunt1975}.

\begin{figure}[!t]
\centering
\includegraphics[width=0.98\columnwidth]{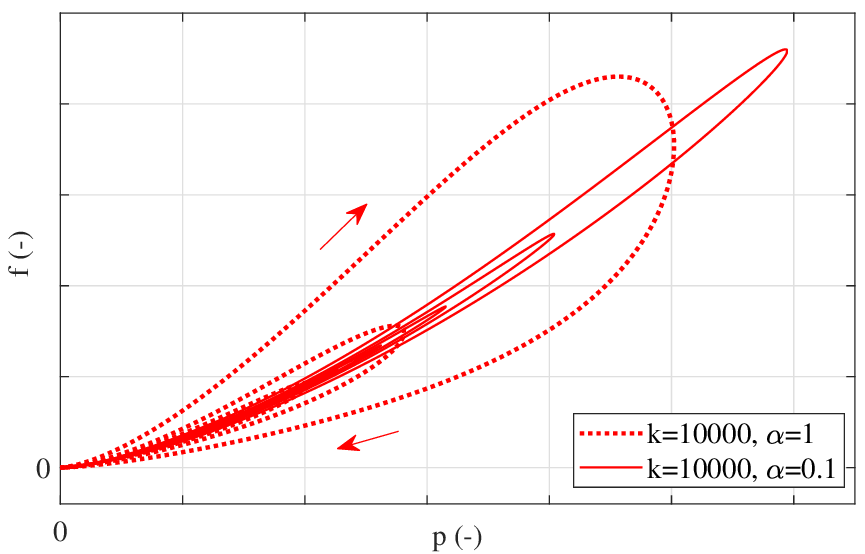}
\caption{Exemplary $(p,f)$ map for different $\alpha$ parameters.}
\label{fig:hystforce}
\end{figure}

One of the important features of the vibro-impact dynamics
\eqref{eq:2:7} is that the contact force is zero at the instants
of impact and separation, the feature which cannot be provided by
the modeling approaches which incorporate a linear damping $\sigma
\dot{x}_2$. Worth noting is also that this is independent of the
velocity magnitude $|\dot{x}_2^i| < \Omega$ at the moment of
contact, while the validity range is limited by some finite $0 <
\Omega < \infty$, cf. \cite{hunt1975}. An illustrative example of
how the contact force \eqref{eq:2:7} develops for a series of
$p(t)$ cycles with a decreasing amplitude is shown in Fig.
\ref{fig:hystforce}.
Here, the arbitrarily assumed stiffness coefficient $k=10000$ is
kept the same, while the compared $\alpha = \{0.1, \, 1\}$ factors
are differing by the order of magnitude. One can recognize the
shaping effect of $\alpha$, that gives rise also to a larger area
of the vibro-impact hysteresis loop and, thus, structural
dissipation at each closed cycle of penetration.

\section{Experimental setup}  
\label{sec:3}

\begin{figure}[!t]
\centering \vspace{1mm}
\includegraphics[width=0.96\columnwidth]{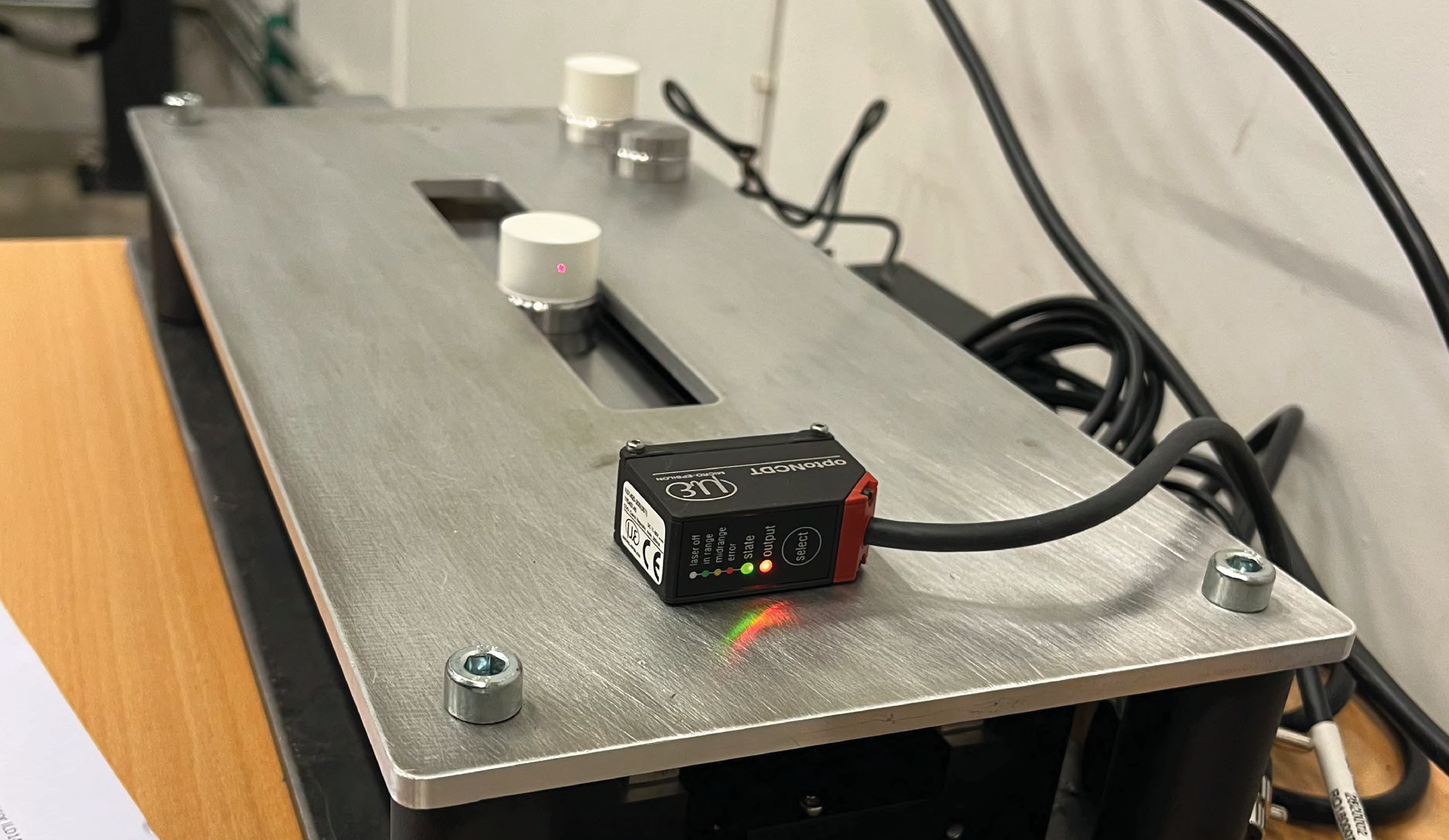}
\caption{Experimental setup of controlled active subsystem with
passive mass put on the flat surface of the moving platform.}
\label{fig:expsetup}
\end{figure}

A tribological experimental setup used in this study is shown
(laboratory view) in Fig. \ref{fig:expsetup}, see
\cite{ruderman2023robust} for details. The linear moving platform,
which is an active subsystem, is arranged under and aligned with
the fixed mechanical frame. The lumped disk, which is a passive
subsystem $m_2$, is guided within the mechanical frame. Note that
the lateral contacts between the disc and the walls of the
frame-slot are minimized by the specially machined side-edging on
the disc. This way the side-contact friction with the walls can be
neglected, and the single essential frictional interface is
associated the disc staying horizontally on the moving platform.
Two disk samples are used: one from the steel with the mass
$m_2'=0.052$ kg, and one from the aluminium with the mass
$m_2''=0.024$ kg. The corresponding Coulomb friction coefficients
are $b' = \mu' m_2' \, g = 0.214$ N and $b'' = \mu'' m_2'' \, g =
0.1106$ N, where $g$ is the gravitational acceleration constant
and $\mu'$, $\mu''$ are the nominal friction coefficients. The
latter are assumed as the known standard values for the
steel-on-steel and aluminium-on-steel pairs, respectively. The
moving platform is actuated by a servo-drive with the stiff
high-precision ball-screw, so that the feedback controlled linear
displacement $x_1(t)$ can be directly commanded. The constant
steady-state velocity, used in one of the following experiments,
is $\dot{x}_1(t)= 0.1$ m/sec. This is the maximal possible value
provided by specification of the BLDC-motor driven ball-screw
stage. The absolute position of the passive disk $x_2(t)$ is
measured by a high-resolution laser sensor with the nominal
repeatability of 8 $\mu \mathrm{m}$. The sampling rate of the data
acquisition real-time board is set to 5 kHz.

\section{Comparison of the modeled and measured response}  
\label{sec:4}

\begin{figure}[!h]
\centering
\includegraphics[width=0.55\columnwidth]{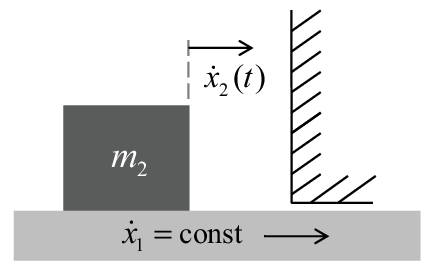}
\caption{Schematic representation of the active-passive pair.}
\label{fig:setup}
\end{figure}

Both modeling frameworks, summarized in sections \ref{sec:2:sub:1}
and \ref{sec:2:sub:2}, are combined to represent the dynamic
behavior of the passive body $m_2$ experiencing an elastic contact
with the fixed frame under the condition $\dot{x}_1 =
\mathrm{const}$. This leads to reduction of
\eqref{eq:2:2}--\eqref{eq:2:4} and, after incorporating
\eqref{eq:2:7}, results in the overall hybrid system dynamics
\begin{eqnarray}
\label{eq:4:1}
  x_1-x_2 &=: & z, \\[2mm]
\label{eq:4:2}
  \dot{x}_1 &=& \mathrm{const}, \\
\label{eq:4:3} \frac{1}{2} \Bigl(1 + \mathrm{sgn} \bigl( x_2 - X_c
\bigr) \Bigr) \, \bigl( x_2 - X_c \bigr)  & =: & p, \\
\label{eq:4:4}
m_2 \ddot{x}_2 - b\,\mathrm{sgn}(\dot{z}) + f(p,t)
&=& 0.
\end{eqnarray}
The threshold value $X_c = \mathrm{const}$ represents the position
of the impact. Following to that, the internal (virtual) state
$p(t)$ constitutes the penetration of the impacting body $m_2$
into an elastic structure of the fixed frame, cf. Fig.
\ref{fig:setup}.

Two different experimental scenarios were conducted and compared
with the modeled behavior. Both are designed to be conclusive and
complementary to each other in evaluation of the modeling
framework \eqref{eq:4:1}--\eqref{eq:4:4}.

\begin{figure}[!h]
\centering
\includegraphics[width=0.98\columnwidth]{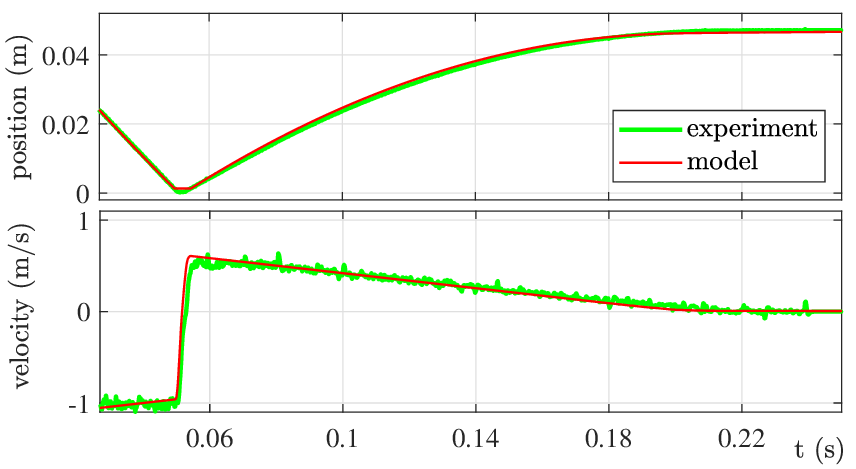}
\caption{Measured and model fitted vibro-impact response.}
\label{fig:impact}
\end{figure}

In the first one, the active mechanical subsystem was in the idle
state, implying $\dot{x}_1 = \mathrm{const} = 0$, and an impulsive
excitation was provided to the passive body via a rubber hammer,
that is often used in the structural modal analysis. For this type
of experiments, the steel disk was taken since it discloses a more
stable translational motion during and after the impact with the
fixed frame, i.e. no tilting-type separation of the horizontal
contact from the supporting platform (i.e. active subsystem
$m_1$). The enforced excitation provided a sufficiently high
relative velocity of the free moving body $m_2$, and allowed for
the values $\dot{x}_2^i \approx 1$ m/sec before the collision. The
measured and model fitted vibro-impact response is exemplary shown
in Fig. \ref{fig:impact}. Note that the experimental relative
velocity $\dot{x}_2$ is obtained by the discrete time
differentiation of the measured position $x_2(t)$ with a
subsequent low-pass filtering.

\begin{figure}[!h]
\centering
\includegraphics[width=0.98\columnwidth]{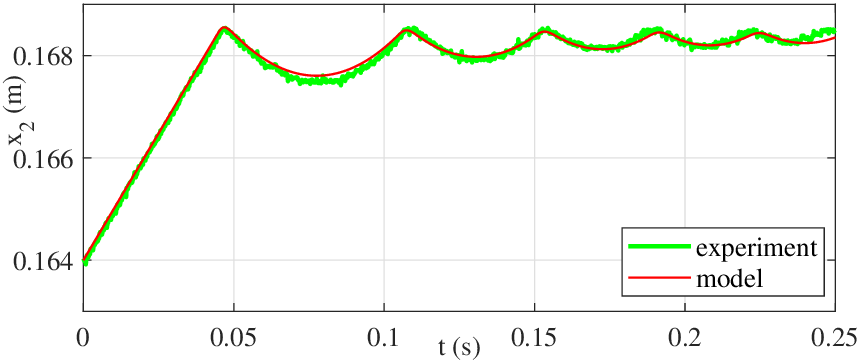}
\caption{Measured and model fitted vibro-impact response.}
\label{fig:fittedvibro}
\end{figure}

The second experimental scenario was designed for the maximal
possible constant velocity of the active subsystem $\dot{x}_1 =
0.1$ m/sec while the passive body $m_2$ came to contact with the
frame and, afterwards, experienced a series of vibro-impact
transitions. The latter occurred due to an interplay of the
repulsive contact force, Coulomb friction damping, and a
continuous motion of the active part of the contact pair, i.e.
$\dot{x}_1 = \mathrm{const} \neq 0$, cf.
\eqref{eq:4:1}--\eqref{eq:4:4}. Here the aluminium disk was used
due to its lower mass and friction coefficient $\mu''$, that leads
in a larger repulsive displacement and, thus, higher number of the
vibro-impact cycles. The measured and model-fitted vibro-impact
response are exemplary shown in Fig. \ref{fig:fittedvibro}. Note
that $t=0$ and $x_2(0)$ are shifted here for the sake of a better
visualization, while $X_c$ parameter is highly sensitive due to
the process noise and uncertain touching point of the laser beam
to the moving disc $m_2$, cf. Fig. \ref{fig:expsetup}.

\section{Conclusions}  
\label{sec:5}

In this paper, an experimental case study was developed, performed
on a dedicated tribological setup, \cite{ruderman2023robust},
while combining the Coulomb based friction coupling
\cite{ruderman2022dynamics} with the nonlinear structural damping
modeling \cite{ruderman2021stiffness} based on \cite{hunt1975}.
The overall hybrid dynamics framework was formulated for the given
mechanical structure, and two specific experimental scenarios were
designed and performed for collecting the motion data of a passive
inertial body. This one was subject to elastic vibro-impact with a
fixed frame and normal frictional interface to an active body --
the moving and supporting platform. It was shown how the nonlinear
mechanisms of the Coulomb friction coupling and the contact
restitution are superimposed and able to adequately capture the
transient system behavior. The demonstrated accord between the
measured and model-predicted displacement response supported this.


\bibliographystyle{IEEEtran}
\bibliography{references}

\end{document}